\documentclass[aps,prl,reprint]{revtex4-2}

\usepackage{graphicx}
\usepackage{dcolumn}
\usepackage{bm}
\usepackage{amsmath}
\usepackage{xcolor}
\usepackage{ulem}
\usepackage{bbm}
\usepackage{dsfont}

\begin{document}

\title{White light interferometry analysis for measuring thin film thickness down to few nanometers}

\author{Victor Ziapkoff}\email{victor.ziapkoff@universite-paris-saclay.fr}

\author{François Boulogne}\email{francois.boulogne@cnrs.fr}

\author{Emmanuelle Rio}\email{emmanuelle.rio@universite-paris-saclay.fr}

\affiliation{Université Paris-Saclay, CNRS, Laboratoire de Physique des Solides, 91405, Orsay, France}

\author{Anniina Salonen}\email{anniina.salonen@espci.fr }
\affiliation{Soft Matter Sciences and Engineering, ESPCI Paris, PSL University, CNRS, Sorbonne Université, 75005 Paris, France}

\date{\today}

\begin{abstract}

We present a practical white-light interferometric method, supported by an open-source Python library \textit{optifik} for automated spectrum-to-thickness deduction, enabling foam film measurements down to a few nanometers. We describe three typical spectral scenarii encountered in this method: spectra exhibiting numerous interference fringes, spectra with a moderate number of peaks, and spectra with only a few identifiable features, providing illustrative examples for each case. We also discuss the main limitations of the technique, including spectral range constraints, the necessity of knowing the refractive index, and the influence of spectral resolution and signal quality. Finally, we demonstrate the application of the method in a time-resolved study of a TTAB (tetradecyltrimethylammonium bromide) foam film undergoing elongation and thinning. This method can be adapted to measure any thin non-opaque layer.

\end{abstract}

\maketitle

\section{Introduction}\label{sec1}

In numerous industrial and scientific contexts, including coatings \cite{Kistler1997}, accurate measurement of the thickness of thin liquid films is a key parameter to characterize the system properties. 
Despite its importance in various situations (process control, material characterisation) measuring widely varying film thicknesses is challenging, and often requires a combination of techniques. 

Foam films provide a particularly illustrative example of such systems. 
They experience a stratification under gravity, leading to a thickness profile increasing from the top to the bottom. 
Additionally, they thin over time with thicknesses ranging from micrometers for young soap films down to a few nanometers before their rupture. Because of the associated interference colours, this leads to fascinating patterns varying in space and time.

As detailed in the article \textit{Chronicles of foam films} by Gochev $\textit{et al.}$ \cite{gochev_chronicles_2016}, soap films have not only been used by artists in various contexts, but have also been studied for centuries by scientists. From a chronological perspective, the earliest film thickness measurements are based on the association between the color of a soap film and interference patterns, a phenomenon first studied by Newton \cite{newton_opticks_1952}. 
Visual measurements are possible down to the last color in the color band corresponding to the first order of constructive interference. 
He notably determined that a foam film with that color has a thickness of approximately 107~nm. 

In the late 19th century (1877-1893), Reinold and Rücker \cite{ReinoldRucker1881,ReinoldRucker1893} investigated the thickness of foam films using electrical resistance and optical interference methods at known and fixed humidity and temperature. Notably, they are the first to report film thicknesses of approximately 11-12~nm.

Later, Johonnot \cite{johonnott_xlviii_1899} employed two experimental methods --- both originally proposed by Michelson --- to measure the thickness of thin films using a monochromatic light source (a sodium vapor lamp with wavelength $\lambda = 589~\mathrm{nm}$). 
The first method is interferometric, while the second is photometric. 
In the interferometric approach, $N$ soap films are placed in one arm of a Michelson interferometer. 
The thickness is obtained by observing the displacement of vertical interference fringes. In the photometric method, the thickness of a single film is inferred from the intensity of reflected light measured at various angles of incidence.

By observing vertically oriented fluorescein films and horizontally stratified potassium oleate films, Perrin proposed a "multiple-layer law" \cite{perrin_stratification_1918}. 
According to this model, stratified films are composed of integer multiples of an elementary layer, with a characteristic thickness of 5.2~nm. 
By counting the number of visible stratification layers, Perrin was able to correlate observed film colors with thickness and thereby confirm the thickness of the elementary layer.

In 1950, Miles, Ross, and Shedlovsky \cite{miles_film_1950} investigated thin film drainage using a horizontal film apparatus illuminated by a fluorescent lamp. 
Building on the earlier observations of Dewar and Lawrence \cite{lawrence_soap_1929}, they measured film thicknesses down to 5-10~nm. 
This early optical approach was later refined by Mysels, Shinoda, and Frankel, who used the intensity of a monochromatic light reflected from specific regions of the film to determine a local film thickness \cite{mysels_soap_1959}.

Advances in thickness measurements accelerated with the development of the thin film pressure balance (TFPB) by Scheludko and Exerowa, based on the apparatus of Derjaguin and Titijevskaya \cite{derjaguin_stability_1953,derjaguin_surface_1957}. 
This technique enabled the precise control of capillary pressure across microscopic horizontal foam films, allowing for the first time the experimental validation of the DLVO theory, independently proposed by Derjaguin and Landau \cite{derjaguin_theory_1941} and by Verwey and Overbeek \cite{verwey_theory_1948}. 
Film thickness was typically obtained \textit{via} microinterferometry. 
This technique, grounded in Scheludko’s renormalization framework \cite{scheludko_ann_1954}, could be performed using either monochromatic or white-light sources. 

Fourier-transform infrared (FTIR) spectroscopy is also used to determine not only the thickness of soap films but also the orientation and phase behavior of hydrocarbon chains within the film \cite{Umemura_1985}. In addition, ellipsometry, which exploits the polarization change of reflected light, is another powerful technique for characterizing thin films \cite{DenEngelsen1974}. 

In the last 15 years, the techniques developed in the past have been updated with modern instrumental developments. The commercialization of white light optical fibers coupled with spectrometers allowed the acquisition of instantaneous and localized spectrum of the reflected light \cite{champougny_influence_2018,champougny2016life,miguet2020stability,pasquet_lifetime_2024,pasquet2022impact}. More recently, a hyperspectral camera has been developed, which allows the acquisition of 254 light spectra on a line \cite{chandran_suja_hyperspectral_2020,shabalina2019rayleigh,etienne-simonetti_hydrodynamic_2024,gros2021marginal}. 
Even more recently, researchers proposed to exploit the whole interference pattern, \textit{i.e.} the position, curvature and number of fringes, at the surface of a bubble to extract the entire thickness maps \cite{miguet2024absolute}.

Beyond visible optics, other techniques such as reflectivity, diffraction, and scattering techniques with X-rays and neutrons have become indispensable for probing the structure of interfaces and films \cite{terriac2007characterization,simister_structure_1992,benattar_x-ray_1992,penfold_neutron_2014,mikhailovskaya_probing_2017, chiappisi_liquid_2024}. Complementary approaches based on lateral conductivity \cite{Sheludko1954_1957} and capacitance \cite{Sonntag1960_1964} are also used for measuring film thicknesses. A comparative analysis of several of these techniques is available in Ref. \cite{scheludko_thin_1967}. 

In the present work, we focus on white-light interferometry. 
The different models at stake are already well-known and documented. 
Here we propose to reuse existing models to develop \textit{optifik}, an open-source Python library implementing the different methods. Section~\ref{sec2} details the derivation of Scheludko's renormalization approach for determining thickness differences between successive interference orders. 
In Section~\ref{sec3}, we identify three typical experimental scenarii and discuss the specific considerations for each. 
Section~\ref{sec4} addresses the limitations and sources of error inherent to the method. 
A time-resolved implementation of the technique is explored in Section~\ref{sec5}, with concluding remarks in Section~\ref{sec6}.

\section{Derivation of film thickness by interferometry}\label{sec2}

When white light is shone perpendicularly on a soap film, some of the incident light $I_0$ is reflected at the first liquid-air interface and some is transmitted through the soap film and reflected at the second liquid-air interface. Another part of this incident light is transmitted through the soap film and may, in turn, be reflected on the second liquid-air interface. The rays, resulting from one or more reflections and transmitted by the first interface, interfere with each other. In summary, the soap film behaves as a Fabry-Perot cavity. The reflected light intensity is denoted $I_r$. The phase shift $\phi$ between two successive rays depends on the film thickness $h$ of refractive index $n$ and writes 
\begin{equation}
    \phi = \frac{2\pi}{\lambda} \delta,
    \label{eq:Phiexpression}
\end{equation}

\noindent where $\delta = 2n(\lambda)h$ is the optical path difference. 
In practice, the refractive index $n$ is wavelength dependent.
The resulting reflected intensity follows  Airy's formula \cite{airy_undulatory_1877}:
\begin{equation}
    I_r = I_0 \frac{\sin^2\left(\frac{\phi}{2}\right)}{\frac{4n^2}{\left(n^2-1\right)^2} + \sin^2\left(\frac{\phi}{2}\right)}.
    \label{eq:Intensityexpression}
\end{equation}

The function $I_r(\phi)$ is periodic. For a given intensity, there are two strategies to extract $\phi$ and then the thickness of the film from Eq.~\ref{eq:Phiexpression}. The first one consists in extracting the period of the function $I_r(1/\lambda)$. The second method relies on inverting the function $I_r(h)$, which is only possible over intervals where this function is monotonic, hence bijective. In the latter approach, the main difficulty lies in identifying the right interference order $p=\delta/\lambda$. 

Destructive interferences correspond to minima of the function $I_r(h)$ and to integer values $p_d$ of the interference order, while half-integer values $p_c$ correspond to constructive interference orders. Between a constructive and a destructive interference order, the intensity curve decreases and $p_c$ and $p_d$ can be written $p_d(m) = m/2$ and $p_c(m) = (m+1)/2$, with $m$ even, and conversely, between a destructive and a constructive order, the intensity curve increases and $p_d(m) = (m+1)/2$ and $p_c(m) = m/2$ with $m$ odd.

\begin{center}
\begin{figure}[h]
    \centering
    \includegraphics[width=0.45\textwidth]{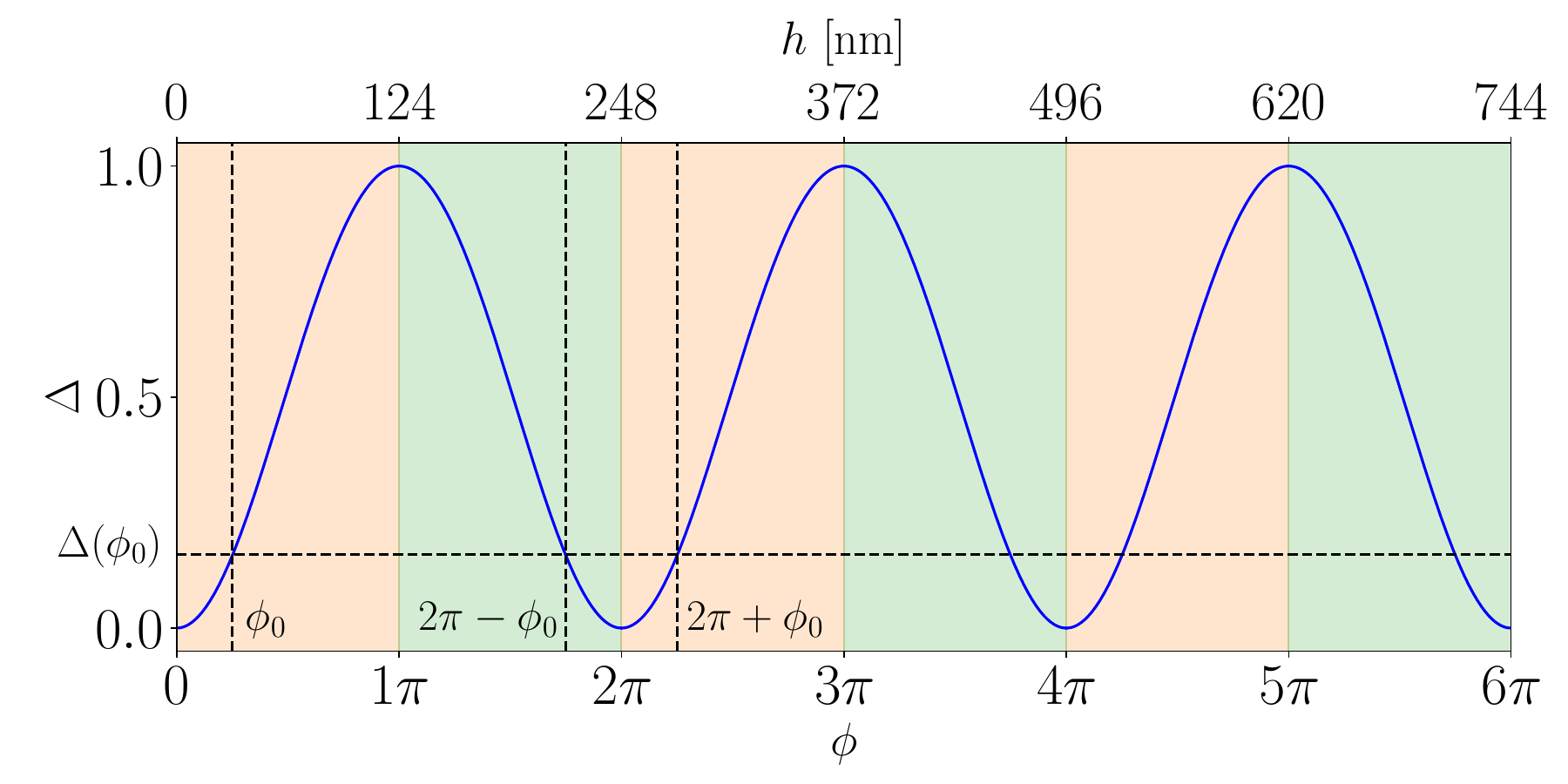}
    \caption{Variation of $\Delta$ as a function of $\phi$ (blue curve). Intervals shown in orange represent regions where the interference order $p$ lies between a destructive interference order $p_d$ and a constructive interference order $p_c$ that satisfy $p_d < p_c$, and \textit{vice versa} for the intervals shown in green. Film thickness using Eq.~\ref{eq:h_expression} with $\lambda = 660$~nm and  $n=1.333$ at three consecutive destructive interference orders $p_d \in [0,1,2]$ and three consecutive constructive interference orders $p_c \in [0.5,1.5,2.5]$ is indicated along the top horizontal axis. Periodicity in the phase $\phi$ and symmetry are show with three black vertical lines at $\phi_0$, $2\pi + \phi_0$, and  $2\pi - \phi_0$ and the horizontal line $\Delta(\phi_0)$.}
    \label{fig2}
\end{figure}
\end{center}

Scheludko \cite{scheludko_thin_1967} proposed a renormalization method to extract film thickness between any pair $(p_c,\,p_d)$ \cite{scheludko_thin_1967}. Let the normalized intensity \( \Delta \) be defined by:
\begin{equation}
\Delta = \frac{I_r - I_{\min}}{I_{\max} - I_{\min}},
\label{eq:DELTA_definition}
\end{equation}
\noindent
where $ I_{\min}$ and $I_{\max}$ are the reflected intensity for destructive and constructive interferences, respectively.
Fig.~\ref{fig2} shows in blue the variation of $\Delta$ as a function of $\phi$. The variation is $2\pi$-periodic and symmetric to $\pi$ modulo [2$\pi$] such that a given value of $\Delta$ corresponds to different values of $\phi = 2k\pi \pm \phi_0$ with $k \in \mathds{N}^*$. This equality is illustrated in Fig.~\ref{fig2} with vertical line at $\phi = \phi_0$ and $\phi = 2\pi \pm \phi_0$ and the horizontal line at $\Delta(\phi_0)$. 
These values of $\phi$ are either on increasing or on decreasing parts of the curve, which is denoted by orange and green colours, respectively, in Fig.~ \ref{fig2}.
The orange regions are found to correspond to interference orders $p$ bounded by a pair $(p_c,\,p_d)$ satisfying $p_d(m) < p_c(m)$ with $m$ even. The phase can then be written $ \phi_m = \phi + m\pi$, which leads to
\[
\phi_m = m\pi + 2\arcsin\left(\sqrt{\frac{\Delta}{1 + (1 - \Delta) \frac{(n^2 - 1)^2}{4n^2}}}\right).
\]

The green regions are associated with interference orders $p$, bounded by a pair $(p_c,\,p_d)$ satisfying $p_c(m) < p_d(m)$ with $m$ odd. 
The phase can be written $\phi_m = (m+1)\pi - \phi$, which yields

\[
\phi_m = (m+1)\pi - 2\arcsin\left(\sqrt{\frac{\Delta}{1 + (1 - \Delta) \frac{(n^2 - 1)^2}{4n^2}}}\right).
\]

\noindent Finally, between any pair $(p_c, p_d)$ which is characterized by a unique $m\in\mathds{N}$, we have

\begin{equation}
\resizebox{6.5cm}{!}{$
h = \frac{\lambda}{2\pi n} \left(C_m\pi + (-1)^m \cdot 
\arcsin\left(\sqrt{\frac{\Delta}{1 + (1 - \Delta) \cdot 
\frac{(n^2 - 1)^2}{4n^2}}}\right)\right),
$}%
\label{eq:h_expression}
\end{equation}

\noindent where
\[
C_m =
\begin{cases}
\frac{m}{2} & \text{if } m \text{ is even}, \\
\frac{m+1}{2} & \text{if } m \text{ is odd}.
\end{cases}
\]

\noindent The film thickness $h$ can thus be directly deduced from the experimental parameter $\Delta$, provided the index $m$ --- which uniquely identifies the pair $(p_c,\,p_d)$ bounding the actual interference order $p$ --- is known. 
In Fig.~\ref{fig2}, the top horizontal axis illustrates the corresponding film thickness obtained from Eq.~\ref{eq:h_expression}, with $\lambda = 660$~nm and $n = 1.333$ for three consecutive destructive interference orders $p_d \in [0,1,2]$ (corresponding to $m \in [0,2,4]$) interleaved with three consecutive constructive interference orders $p_c \in [1/2,3/2,5/2]$ (corresponding to $m \in [1,3,5]$) are displayed.

\section{Foam film thickness determination from a single spectrum}\label{sec3}

\subsection{Experimental procedure}\label{31}

\begin{center}
\begin{figure}[h]
    \centering
    \includegraphics[width=\columnwidth]{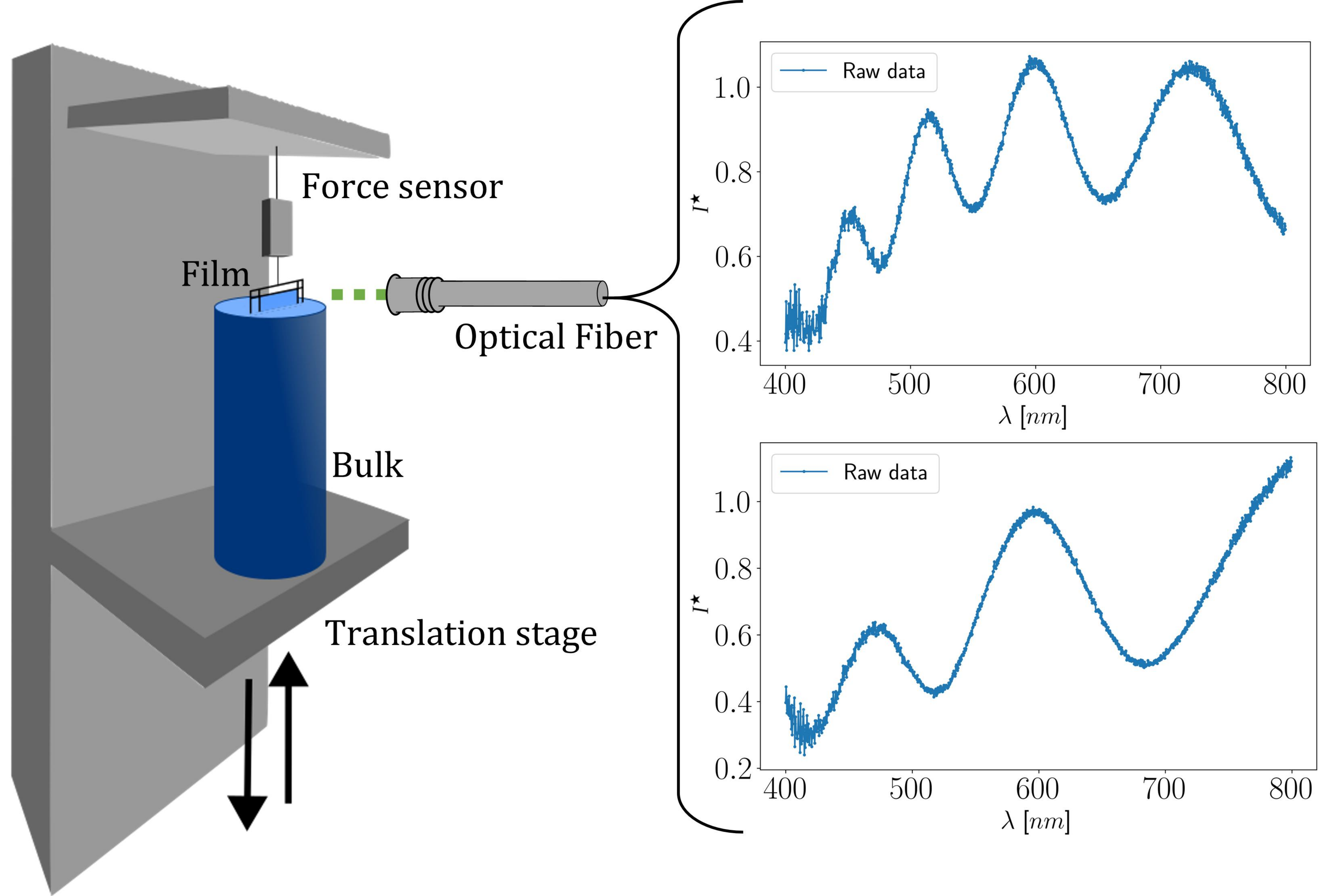} 
    \caption{Set-up scheme with two examples of typical spectra with the wavelength  $\lambda \in [400, 800]~$nm.}
    \label{fig3}
\end{figure}
\end{center}

To illustrate the thickness measurement method, we use the example of a draining vertical soap film. 
Fig.~\ref{fig3} shows a scheme of our setup  \cite{saulnier_study_2014,champougny_influence_2018}. The procedure to generate soap films consists in dipping a 3D-printed rectangular frame of surface $10$~cm $\times$ $1.7$~cm in a circular reservoir of diameter $2.8$~cm and depth $9.7$~cm containing an aqueous solution of recrystallized  TTAB (tetradecyltrimethylammonium bromide --- Sigma Aldrich)  at a concentration of $1.18~$g.L$^{-1}$ (critical micellar concentration). 
Since the refractive index $n(\lambda)$ is wavelength-dependent, we measured its value using a refractometer (Abbemat MW, Anton Paar) for aqueous TTAB solutions at a concentration of  1.2~g.L$^{-1}$ for $\lambda \in $ $[481.3, \ 513.1, \ 589.3, \ 656.2]$ nm and temperature $T = 20 \ ^\circ$C. 
The measurements are  fitted with Cauchy's law, $n(\lambda) = A + B/\lambda^2$ to $A = 1.324$ and $B=3012$ nm$^{2}$, to extract an $\textit{ad hoc}$ function for $n(\lambda)$. 
We note that the addition of TTAB in water has a negligible effect on the refractive index $n(\lambda)$ \cite{Schiebener1990}.

A sub-frame --- two vertical and one horizontal threads --- made with nylon fiber with a diameter of $150~\mu$m is glued at $ 2~$mm from the main frame. The main frame is connected to a force sensor (HBM, 5g) that can detect film rupture. 
The initial position of the manually adjustable translation plate is set so that the horizontal nylon fiber is nearly in contact with the bulk. 
To make the film, the translation plate moves downward at a constant velocity $V = 1~$mm.s$^{-1}$ thanks to a motorized linear stage (Newport UTS150CC) coupled to a motion controller (Newport SMC100CC).
The drainage dynamics of such vertical films are not yet fully understood. 
Recent developments show the primordial role of the marginal regeneration, which leads to thin zones in the vicinity of the lateral vertical meniscus, in which the liquid flux is directed to the top of the film \cite{mysels_interface_1966,seiwert2017velocity}. Monier \textit{et al.} \cite{monier2024self} 
present the most recent experiments enlightening this question. 
Understanding this dynamics is not the primary objective of this study. 
Instead, we focus on measuring the thickness at the top of the film, near the horizontal meniscus --- a region known to thin continuously over time until the film ruptures.

A horizontal optical fiber (IDIL, France) is placed at the very top of the film, at the focal plane of the optical fiber lens. 
The reflected light spectrum is collected by six optical fibers (IDIL, France) surrounding the first one and its spectrum is measured by a spectrometer (NanoCalc 2000 VIS/NIR, Ocean Optics) in the wavelength range $[400, 800]~$nm with a full width at half maximum (FWHM) of 1.5 nm. 
The sensitivity is maximal in the range $[450, 800]~$nm. 
The normalized reflected intensity $I^\star$ is calculated by the expression :

\begin{equation}
I^\star = \frac{I_\text{raw} - I_\text{noise}}{I_\text{max} - I_\text{noise}},
\end{equation}

\noindent where $I_\text{raw}$ is the raw reflected intensity, $I_\text{noise}$ is the intensity when the emitted white light by the optical fiber is switched off, and $I_\text{max}$ is the intensity reflected by a silicon wafer.

Fig.~\ref{fig3} represents two spectra in the range $[400, 800]~$nm at two different times $t$. It is clear from these examples that the signal-to-noise ratio (SNR) is poor in the range $[400, 450]~$nm. In the following, the range of wavelengths $\lambda$ will thus be restricted to $[450, 800]~$nm. 
Additionally, the spectrum $I^\star$ does not resemble the theoretical spectrum $I_r$. A pseudo-periodicity is recovered, but the amplitudes of the peak intensities are not uniform, in contrast to the idealized case shown in Fig.~\ref{fig2}. 
Although such variations can theoretically be partly explained by the wavelength-dependent refractive index $n(\lambda)$, they are significantly amplified in experimental measurements. 
This enhanced variability arises from the high sensitivity of the setup to the alignment between the optical fiber and the soap film, as well as from the intrinsic transparency of the film, which reduces intensity resolution. 
In this context, $I^\star$ denotes the normalized intensity measured experimentally, while $\Delta$ in Section \ref{sec2} refers to the theoretical normalized intensity reflected by a soap film. 
Nevertheless, sufficient contrast remains to enable a reliable film thickness determination via peak detection, provided a second normalization is performed to take into account the observed minima and maxima. 
We note that spectra obtained from solid films or films deposited on solid substrates would provide improved overall spectral quality.

\subsection{Python library: \textit{optifik}}

The methods described in the forthcoming sections are implemented in the open-source Python library \textit{optifik}, available on both GitHub  \cite{optifik_github} and the Python Package Index (PyPI)  \cite{optifik_librariesio}. 
The article described the library as implemented in version 0.4.3.
The source code is distributed under the GNU General Public License v3 (GPLv3). 
The library builds upon the scientific Python stack, notably relying on numpy \cite{numpy}, scipy \cite{scipy}, scikit-learn \cite{scikit-learn}, and matplotlib \cite{matplotlib} for numerical computation, optimization, machine learning, and visualization. 
Comprehensive documentation  \cite{optifik_doc}, which includes installation instructions, API reference, and illustrative Jupyter notebooks, is provided. 
To ensure reliability and facilitate future development, the library is supported by a suite of unit tests. 
These tests are based on theoretical spectra generated from Eq.~\ref{eq:Intensityexpression} to ensure the accuracy and precision of thickness measurements, as well as experimental records to ensure the operation on realistic data with limited resolution, noise, and intensity variations.
Both the test suite and documentation are automatically validated using a continuous integration (CI) pipeline upon each new code commit.

\subsection{Peak detection}\label{sub31}

The positions of the peaks --- corresponding to constructive and destructive interferences --- are directly related to film thickness. To detect the peak positions, the experimental intensity data $I^\star(\lambda)$ are preprocessed using the $\textit{optifik.analysis}$ module of the $\textit{optifik}$ library using the following procedure. 
First, the spectrum is smoothed with the $\textit{smooth\_intensities}$ function, which applies a third-order Savitzky–Golay filter with a window size of 11 points by default. 
The $\textit{find\_peaks}$ function is then applied to the smoothed spectrum.
This function requires the specification of a prominence parameter $P$, which strongly depends on the experimental setup and the spectral quality. Several values of $P$ were tested, and the most reliable results were obtained with $P = 0.018$. 
The final result of this procedure is the identification of the number and position of the spectral peaks.

In the following section, we present various methods for extracting the film thickness from such spectra. 

\subsection{Thick films: Fast Fourier Transform (FFT) Method}

\begin{center}
\begin{figure}[h]
    \centering
    \includegraphics[width=\columnwidth]{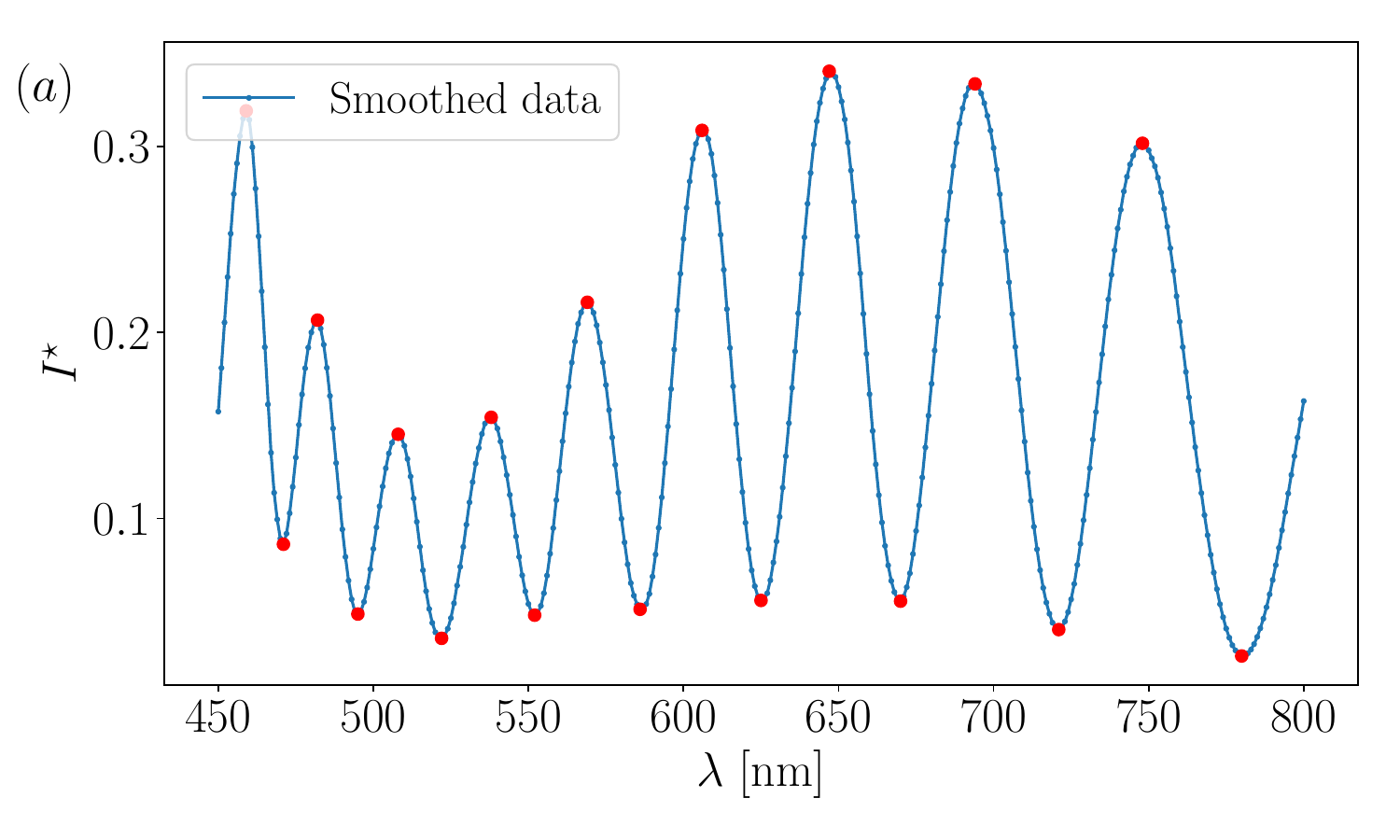} 
    \includegraphics[width=\columnwidth]{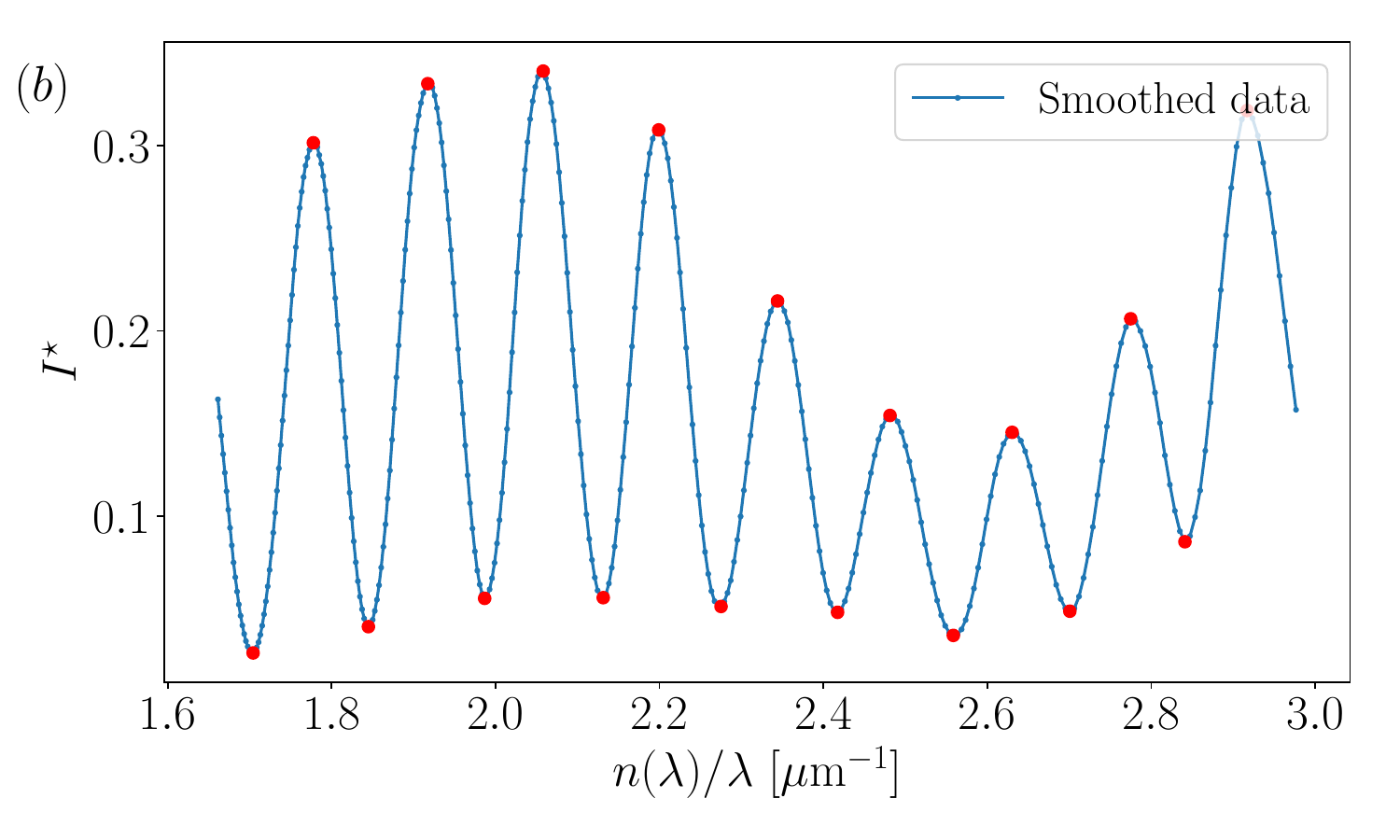} 
    \includegraphics[width=\columnwidth]{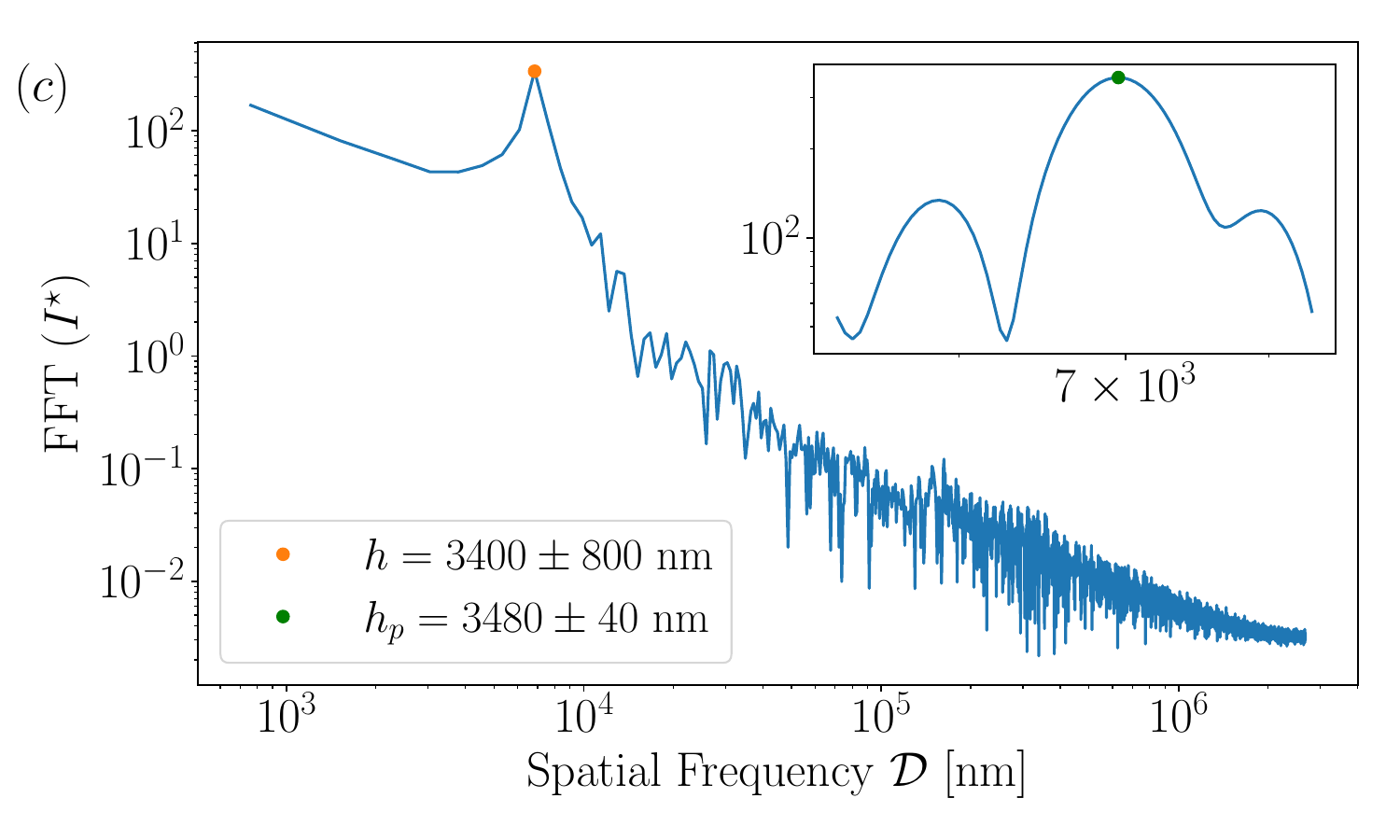} 
    \caption{Smoothed data spectrum in range $[450, 800]~$nm with 18 red dots representing peak detection: ($a$) in the $I^\star(\lambda)$ representation, and ($b$) in the $I^\star(n(\lambda)/\lambda)$ representation. ($c$) Fast Fourier Transform (FFT) on $I^\star (n(\lambda)/\lambda)$. The orange dot represents the dominant spatial frequency $\mathcal{D^\star} = 2h$ with $h$ the film thickness. The inset shows the same FFT spectrum with zero-padding. The green dot represents peak detection with an accuracy improved by a factor 20.}
    \label{fig4}
\end{figure}
\end{center}

When the spectrum contains a large number of peaks, the most suitable method is to make an FFT.
Fig.~\ref{fig4}~($a$) shows a spectrum in which 18 interference peaks are identified after applying the pre-processing procedure described in Subsection~\ref{sub31} on the raw data $I^\star (\lambda)$. 
According to Eq.~\ref{eq:Intensityexpression}, and considering the wavelength-dependent refractive index $n(\lambda)$, $I^\star$ is periodic as a function of $\phi = 4\pi hn(\lambda)/\lambda$.
To extract the period as stated in Section~\ref{sec2}, the data are thus converted from $I^\star (\lambda)$ to $I^\star (n(\lambda)/\lambda)$. 
Fig.~\ref{fig4}~($b$) displays the 18 interference peaks in the $I^\star (n(\lambda)/\lambda)$ representation.

An FFT is then applied to $I^\star (n(\lambda)/\lambda)$ using the $\textit{optifik.fft}$ module of $\textit{optifik}$, as shown in Fig.~\ref{fig4}~($c$). The FFT identifies the dominant spatial frequency, denoted $\mathcal{D}^\star$, present in the spectrum $I^\star (n(\lambda)/\lambda)$. Since interference peaks are approximately regularly spaced, the position of the maximum in the FFT spectrum represents $\mathcal{D}^\star$ highlighted as an orange point in Fig.~\ref{fig4}~($c$). The thickness of the film is therefore directly related to this peak by,

\begin{equation}
h = \frac{\mathcal{D^\star}}{2} = 3400\pm 800~\text{nm}.
\end{equation}

The uncertainty is determined from the step size in the Fourier space, which is set by the length of the signal, $\textit{e.g.}$ the wavelength range. It is possible to reduce this uncertainty by applying a zero-padding to the data. This technique consists in adding neutral data points $I^\star=0$ to extend the data range. This reduces the frequency step in the FFT domain and helps to narrow down the peak detection. The inset in Fig.~\ref{fig4}~($c$) represents the FFT spectrum with a more accurate peak highlighted as a green point, associated with an uncertainty reduced by a factor 20, leading to a thickness,

\begin{equation}
h_p = \frac{\mathcal{D}^\star_p}{2} = 3480\pm 40~\text{nm}.
\end{equation}

The maximum film thickness for which the FFT method remains applicable is constrained by the spectral resolution of the spectrometer $\Delta\lambda=0.33~$nm. 
The upper limit of the measurable optical path difference is determined by the Nyquist–Shannon sampling theorem \cite{shannont_communication_1949,whittaker_xviiifunctions_1915}. According to this theorem, the minimum resolvable optical spacing corresponds to a min-max spacing of $\Delta\lambda=0.33~$nm, $\textit{e.g.}$, a distance of $2\Delta\lambda=0.66~$nm between two successive maxima or minima. 
Applying an FFT to the spectrum yields a maximum optical path length of $230~\mu$m, corresponding to a maximum thickness $h_{\text{max}} = 115~\mu$m.
Note that there can be additional limitations, which depend on the specific set-up.  For example, due to the finite spectral resolution, a decrease in interference amplitude is expected with increasing thickness.

On the other hand, when fewer peaks are detected, fewer periods are present in the spectrum $I^\star (n(\lambda)/\lambda)$, and the FFT method is less accurate. For the range of film thickness $h\in[10^3,\, 10^4]$ nm, we have compared the thickness estimated with the FFT method applied on a theoretical spectrum with the corresponding theoretical film thickness $h$ (Eq.\ref{eq:Intensityexpression}). 
Considering arbitrarily that an error \(\delta h/h = 6\,\%\) is acceptable sets a lower thickness limit of $h = 2.8~\mu$m, corresponding to approximately 15 peaks, below which we use another method (see Section~\ref{subsec33}).

\subsection{Intermediate thickness: Peak fitting}\label{subsec33}

\begin{center}
\begin{figure}[h]
    \centering
    \includegraphics[width=\columnwidth]{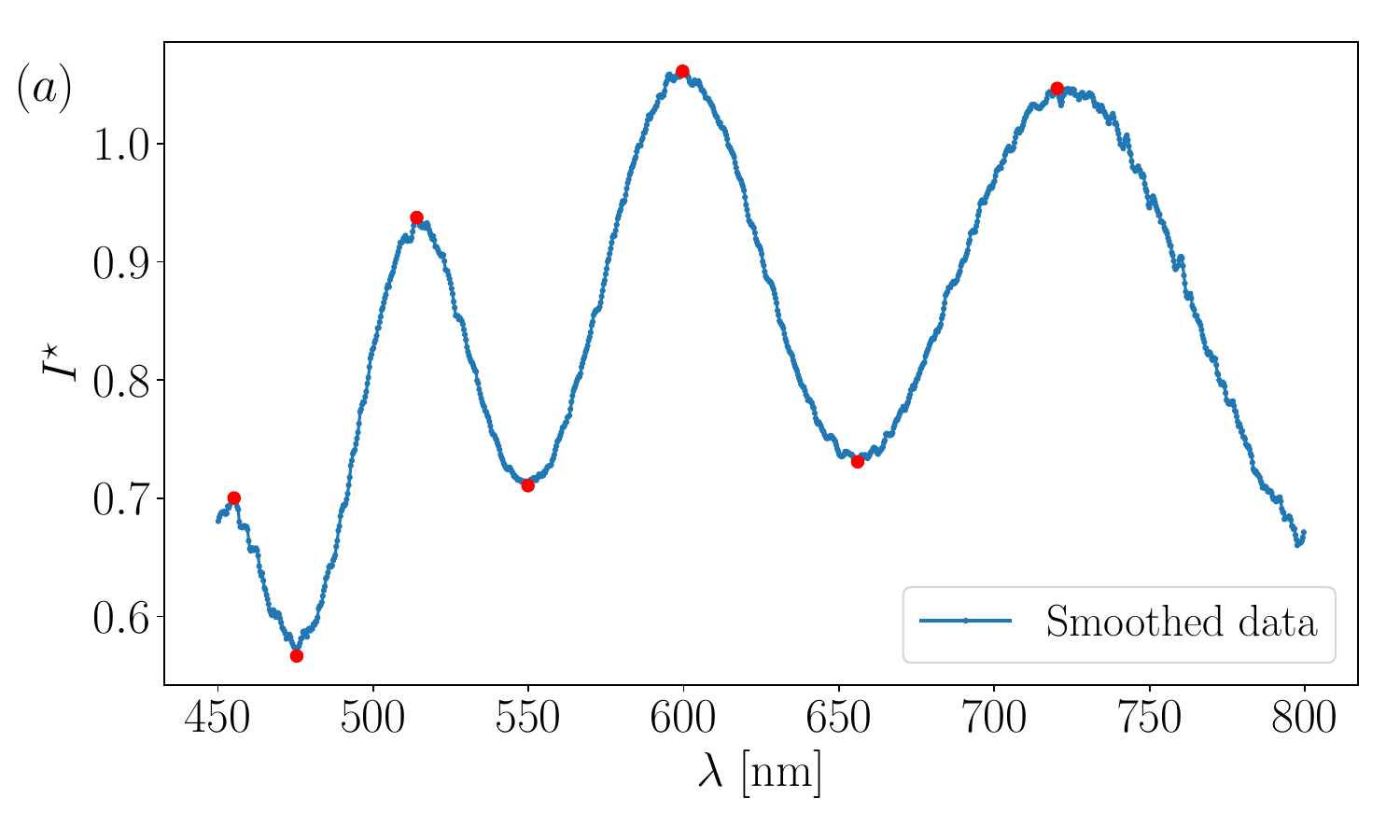} 
    \includegraphics[width=\columnwidth]{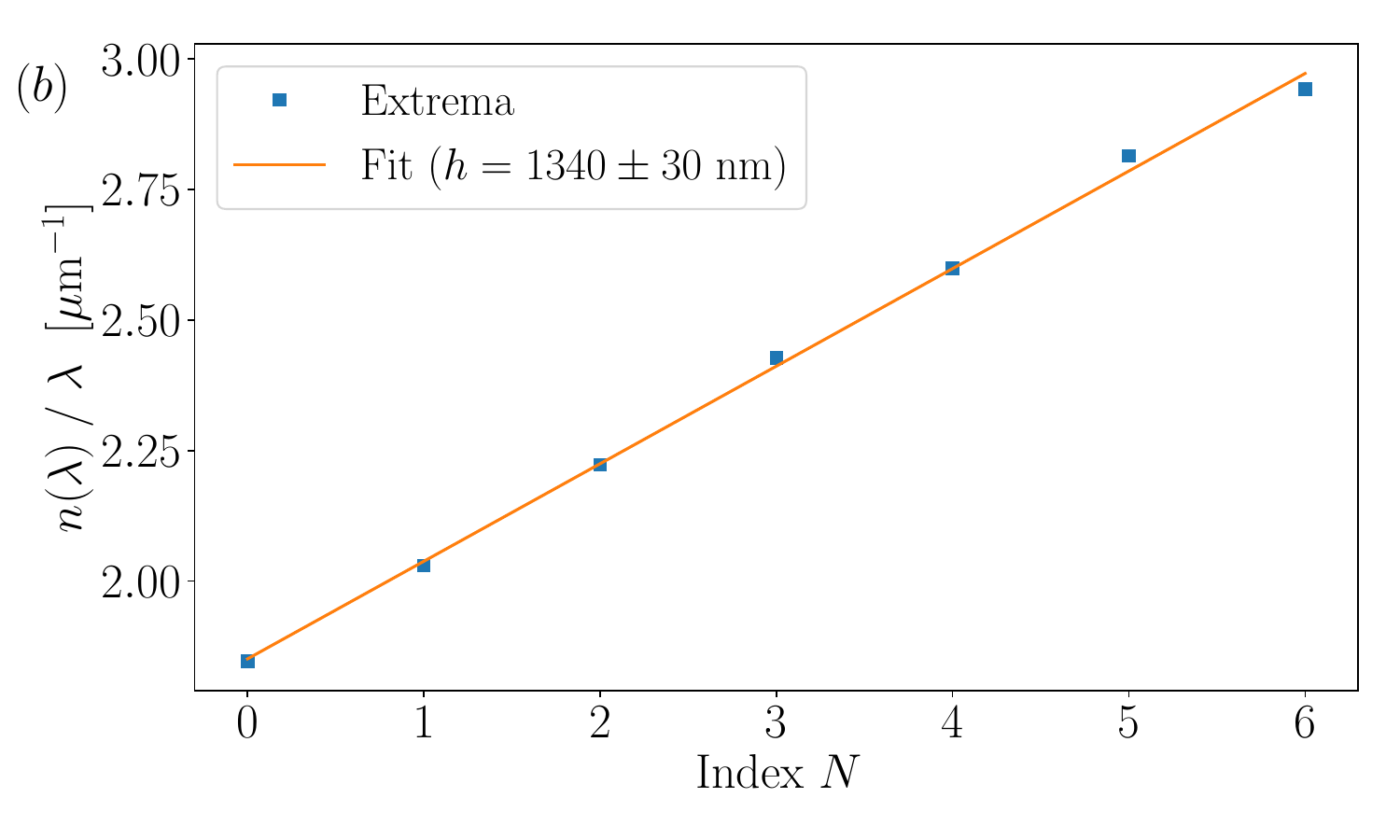} 
    \caption{($a$) Smoothed data spectrum in the range $[450, 800]$~nm with 7 red dots representing peak detection. ($b$) Least squares linear fit of $n(\lambda)/\lambda$ as a function of $N$ giving the slope $1/4h$. For this case, $h = 1340 \pm 30$~nm.}
    \label{fig5}
\end{figure} 
\end{center}

When the spectrum contains only a few peaks, the FFT method becomes less reliable. Fig.~\ref{fig5} ($a$) shows a spectrum with six peaks detected on the smoothed data after the pre-processing. For each peak, the film thickness is related to the interference order $p$, the refractive index $n(\lambda)$, and the wavelength $\lambda$ by the following expression: 

\begin{equation}
    h = \frac{1}{2} \frac{p}{n(\lambda)/\lambda}.
    \label{eq:thickness_InterfrenceOrder}
\end{equation}

By indexing each extremum on the spectrum by $N \in \{0, \ldots, 6\}$, starting with $N = 0$ for the rightmost red dot shown in Fig.~\ref{fig5} ($a$), there is an affine relationship between the interference order $p$ and the index $N$, such that,

\begin{equation}
p=\frac{N + a}{2},
\label{eq:InterferenceOrder}
\end{equation}

\noindent where $a\in \mathds{N}$ is an arbitrary integer denoting the fact that we only know the evolution of the interference order $p$ and not its absolute value. Nevertheless, combining Eq. \ref{eq:thickness_InterfrenceOrder} and \ref{eq:InterferenceOrder}, shows that independently of the value of $a$, $n(\lambda)/\lambda$ varies linearly with $p$ with a slope $h/4$. 

In Fig.~\ref{fig5} ($b$), using the $\textit{optifik.minmax}$ module of $\textit{optifik}$, we plot $n(\lambda)/\lambda$ as a function of $N$. The experimental value of the slope $\alpha$ is determined by applying a least squares linear regression to the data. For this case, the least squares regression gives the thickness at $h = 1340 \pm 30$~nm, whereas the FFT without zero-padding method returns a lower value of $h = 1100 \pm 800$~nm, representing a relative error of about 15~\% due to the loss of accuracy of the FFT method. Additionally, the uncertainty cannot be reduced by zero-padding the FFT, because the peak remains too broad in the spectral domain, whereas the uncertainty from the least squares fit corresponds to the residual standard error of the linear model.

\begin{center}
\begin{figure}[h]
    \centering
    \includegraphics[width=\columnwidth]{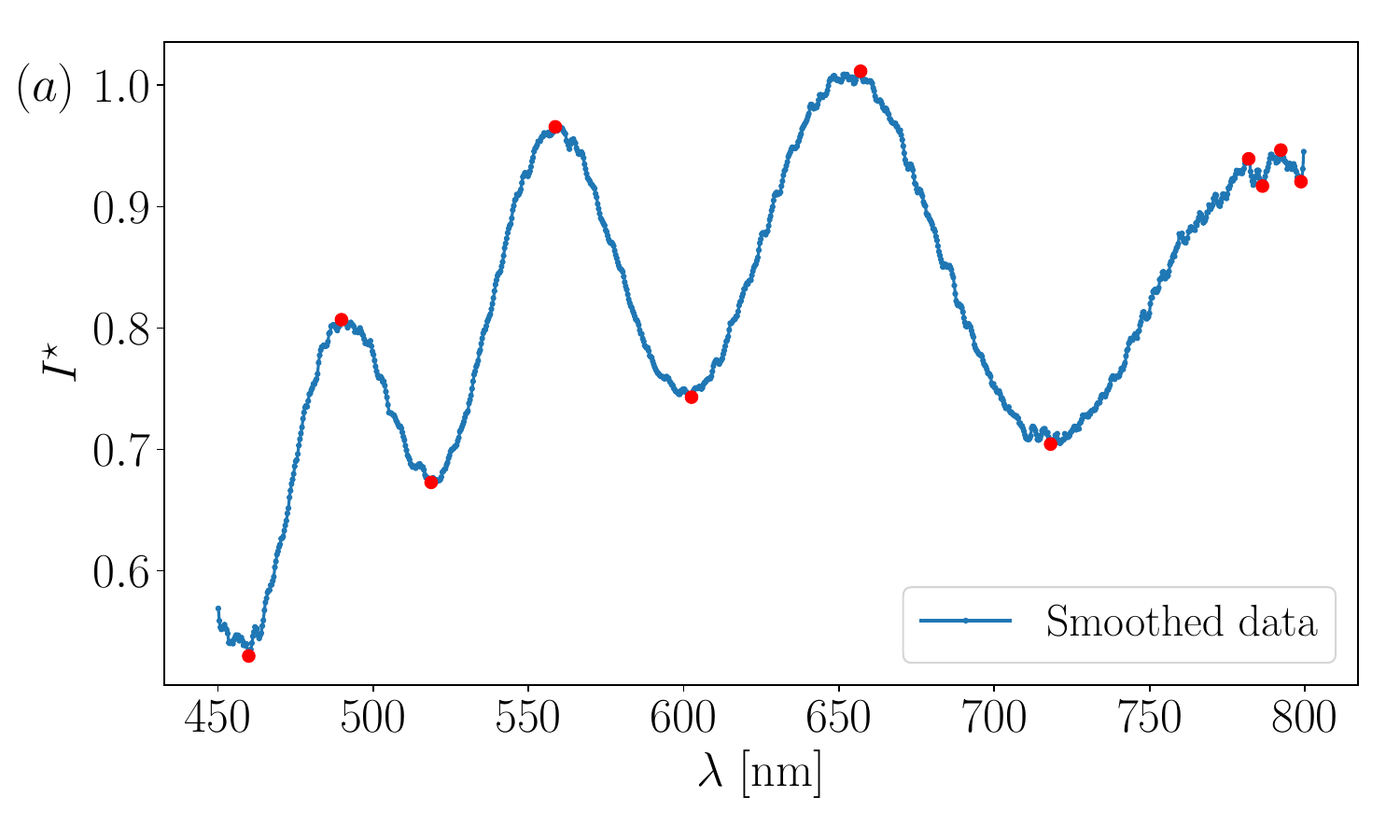} 
    \includegraphics[width=\columnwidth]{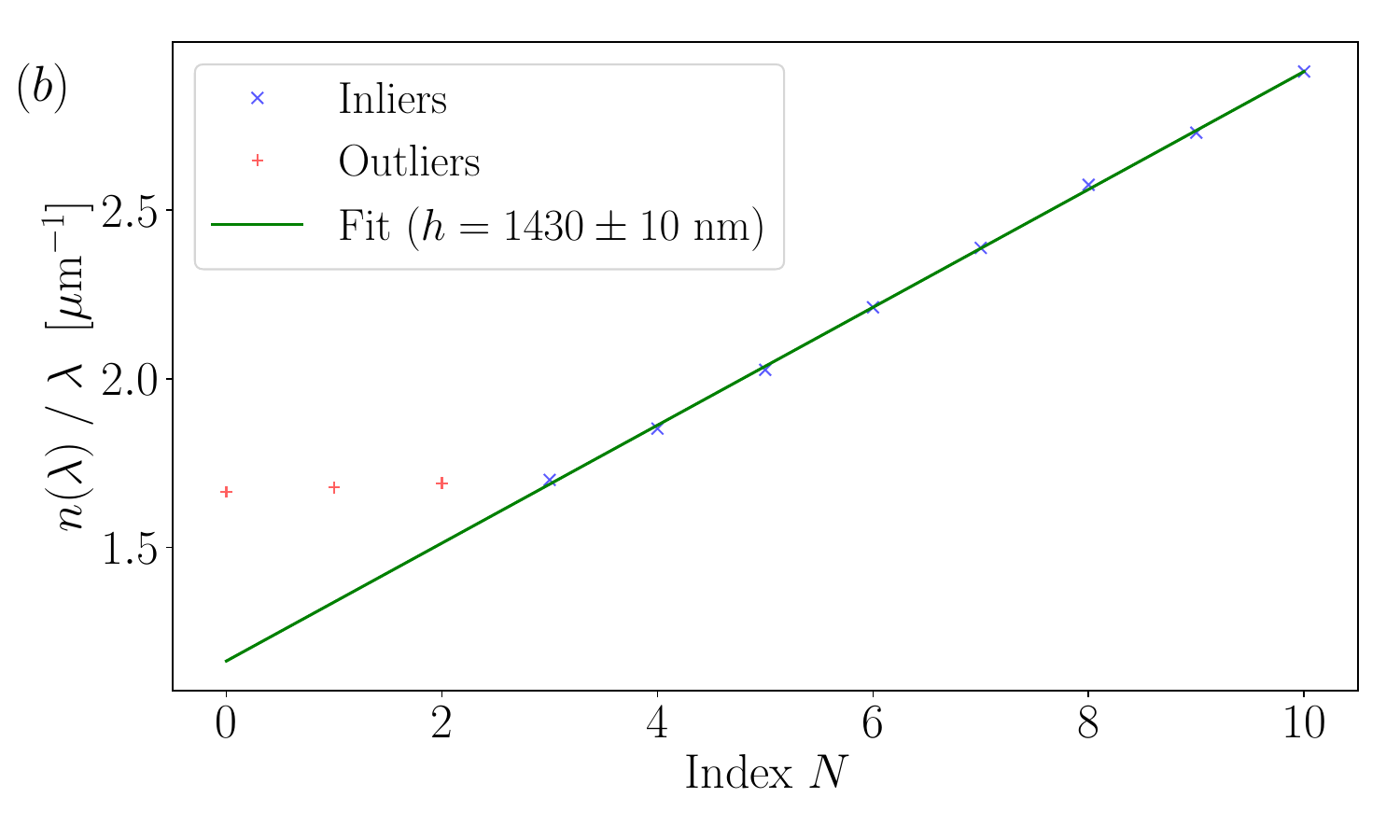} 
    \caption{($a$) Smoothed data spectrum in range $[450, 800]$~nm with 11 red dots representing peak detection. ($b$) Least squares linear fit of $n(\lambda)/\lambda$ as a function of $N$ of the subgroup $N \in \{3, \ldots, 10\}$. For this case, $h = 1430 \pm 10$~nm.}
    \label{fig9}
\end{figure}
\end{center}

The peak detection method may occasionally fail to detect peaks or may detect an excessive number of them. In such cases, applying a least squares linear regression to the entire dataset can result in significant errors. As shown in Fig.~\ref{fig9} ($a$), 11 peaks are detected, but errors are visible on the rightmost data points.

To mitigate such failure, we use the RANSAC (RANdom SAmple Consensus) model, which performs linear fitting on randomly selected subsets of data points. These subsets can be the entire dataset or smaller groups. Each subset is evaluated using a least squares linear fit, and the group yielding the minimal fitting error is identified as the inlier set. The corresponding linear regression provides the estimated slope. Data points not belonging to this group are classified as outliers and are excluded from the final fit. The final fit is represented in Fig.~\ref{fig9} ($b$). The uncertainty from the RANSAC fit corresponds to the residual standard error of the linear model computed for the inlier group. The RANSAC method is not infallible and, when detection fails significantly, the resulting thickness estimate may become unreliable. In such cases, inspecting the spectrum and adjusting the parameters described in Subsection~\ref{31} can be helpful. 

Because the thickness is determined by a linear fit with as many data points as extrema detected, the fewer the extrema, the less effective the calculation of film thickness. 
We observe experimentally that five detected peaks are the lower boundary. It corresponds to a soap film thickness of about 800 nm. In Sec.~\ref{subsec34}, we propose a method to measure smaller thicknesses.

\subsection{Small thickness: Scheludko renormalization}\label{subsec34}

\begin{center}
\begin{figure}[h]
    \centering
    \includegraphics[width=\columnwidth]{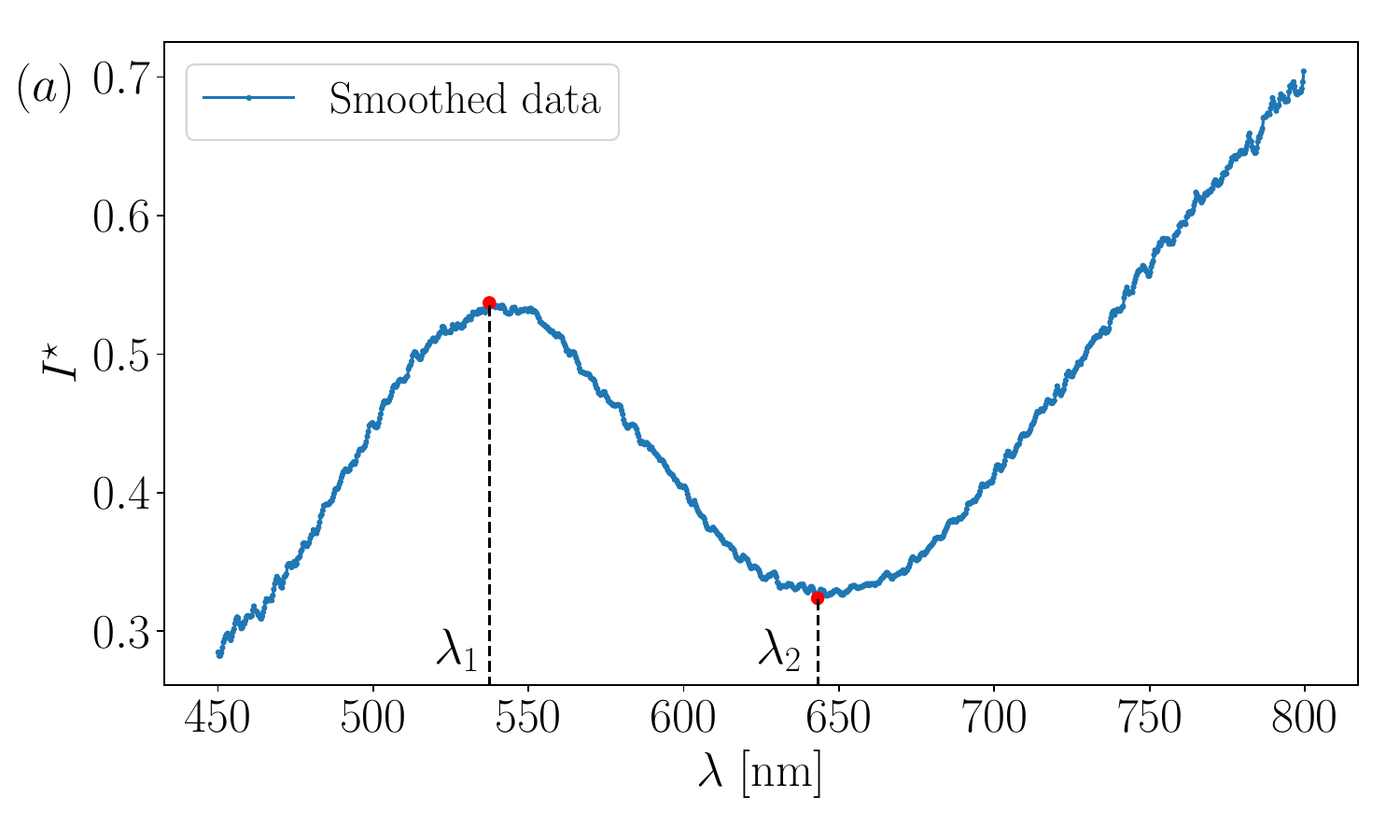}
    \includegraphics[width=\columnwidth]{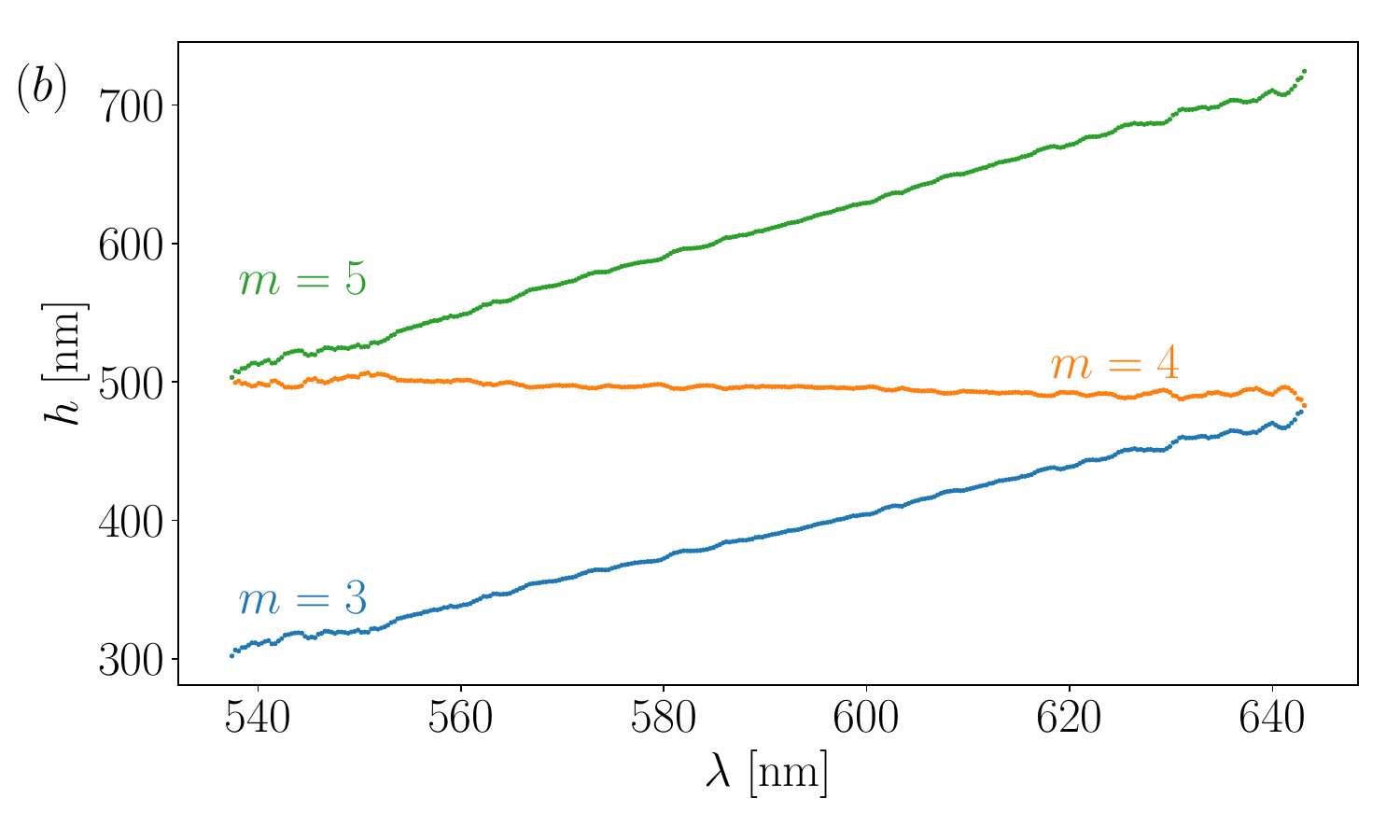}
    \includegraphics[width=\columnwidth]{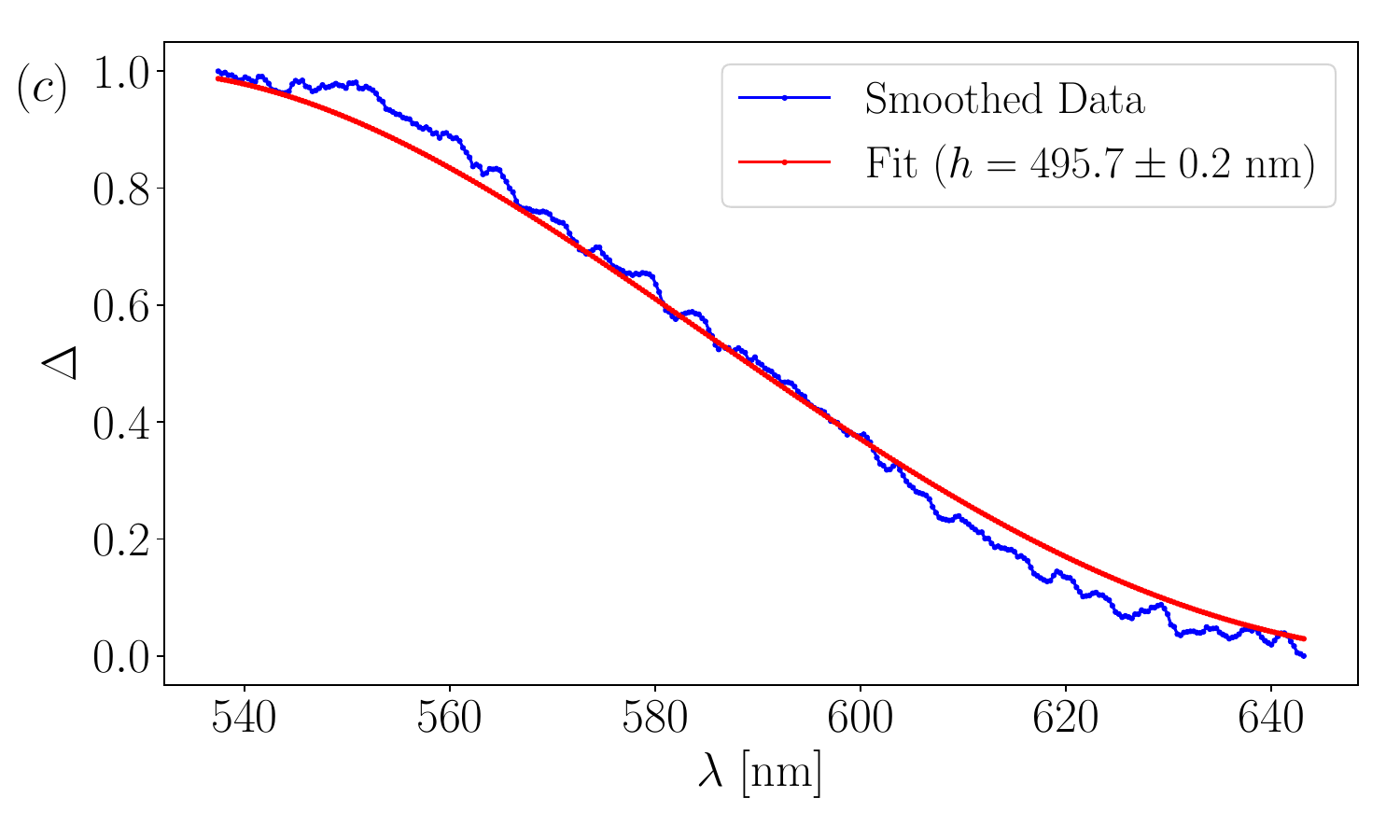}
    \caption{($a$) Smoothed spectral data over the range $[450, 800]$~nm, with two red dots indicating detected peaks. ($b$) Calculation of $h(\lambda)$ within the peak regions using Eq.~\ref{eq:h_expression} for $m = [3, 5]$. ($c$) Calculation of $\Delta(\lambda)$ in the peak range for the smoothed data.
    The red curve corresponds to the $\Delta(h)$ fit obtained via Eq.~\ref{eq:h_expression}, with $h^*$ as the initial guess.
}
    \label{fig6}
\end{figure}
\end{center}

When four or fewer peaks are detected, the Scheludko renormalization method, described in Section \ref{sec2}, becomes effective. 
In such cases, the wavelength difference between two consecutive interference orders $p = m/2$ and $p = (m+1)/2$ is sufficiently large to allow for a reasonably accurate calculation of $\Delta(I^\star)$ \cite{Gaillard2016}. Since the calculation of $\Delta(I^\star)$ necessitates the values of the extrema of $I^\star$, at least one maximum and one minimum are required for its evaluation. An example of such a spectrum is shown in Fig.~\ref{fig6}. In this particular example, to calculate $\Delta$, we use Eq. \ref{eq:DELTA_definition} with $I_{\text{min}} = I^\star(\lambda_2)$ and $I_{\text{max}} = I^\star(\lambda_1)$. This exemplified that $I_{\text{min}}$ and $I_{\text{max}}$ depend on the specific spectrum and order, since $I^\star$ is actually larger for higher wavelength.

Since only a monotonic branch of $I(\lambda)$ is of interest, our strategy is thus to crop the smoothed data outside the two wavelengths, $\lambda_1$ and $\lambda_2$, corresponding to the two rightmost extrema. 
Then, for each wavelength $\lambda \in [\lambda_1, \lambda_2]$, the film thickness $h$ is computed using Eq.~\ref{eq:h_expression} for each interference order $m \in \{0, \ldots, 6\}$. The thickness is calculated using the $\textit{optifik.scheludko}$ module of $\textit{optifik}$ and plotted in Fig.~\ref{fig6}~($b$) for $m = [3, 5]$. 

For the right $m$, all wavelengths should yield the same thickness. It seems to be true for $m=4$ in Fig.~\ref{fig6}~($b$). 
To quantify this observation,  we compute the difference between the maximum and minimum thickness values for each value of $m$, $\textit{e.g.}$, $\min_m[\max (h_m(\lambda)) - \min(h_m(\lambda))]$. 
The value of $m$, for which this quantity is minimized, is the right $m$ denoted $m^*$, and the corresponding mean thickness is defined as $h^*$.
In Fig.~\ref{fig6}~($b$), $m^* = 4$, which can thus be identified as the right interference order. 
We use $h^*$ as an initial guess and perform a fit of $\Delta(h)$ by Eq. \ref{eq:h_expression} to refine the estimate of the film thickness. 
The obtained expression for data is plotted in Fig.~\ref{fig6}~($c$) in red. The uncertainty associated with the fit corresponds to the residual standard error of the single adjustable parameter $h$.

Note that performing the same procedure for the third and the second rightmost extrema, if they exist, shall lead to a new $m = m^* + 1$ but the same $h^*$. 
A difference in the determination of $h^*$  could indicate that the spectrum differs from theoretical cases due to experimental factors.

If the spectrum has a single extremum, specifically a maximum located at wavelength $\lambda_0$, then plotting Eq.~\ref{eq:Intensityexpression} over the range $\lambda \in [450, 800]$~nm leads to two possible cases. 
The maximum may correspond either to an interference order $p_d = 3/2$ ($m=3$), in which case the film thickness cannot be determined from the spectrum, or to $p_d = 1/2$ ($m=1$). In the latter situation, the minimum intensity $I_\text{min}$, extracted from the spectrum acquired without the film, can be used to compute $\Delta(\lambda)$ via Eq.~\ref{eq:h_expression} within the interval $\lambda \in [\lambda_0,\,800]$~nm. 
This enables the measurement of film thicknesses in the range $h \in [84,\,150]$~nm, where the limiting values correspond to interference maxima located at the boundaries of the available spectral range, $\textit{e.g.}$ $\lambda_0 = 450$~nm and $\lambda_0 = 800$~nm, respectively.

Finally, if the spectrum contains only one minimum or no detectable extrema, the calculation cannot be performed. The following section discusses the limitations of this method in more detail.

\section{Limitations in the different methods of analysis}\label{sec4}

\begin{center}
\begin{figure}[h]
    \centering
    \includegraphics[width=\columnwidth]{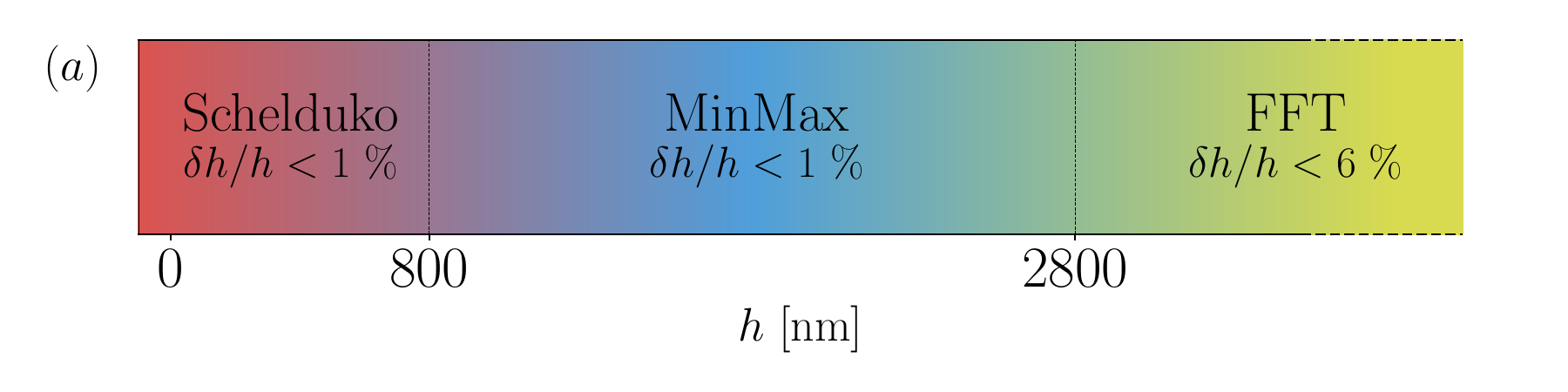} 
    \includegraphics[width=\columnwidth]{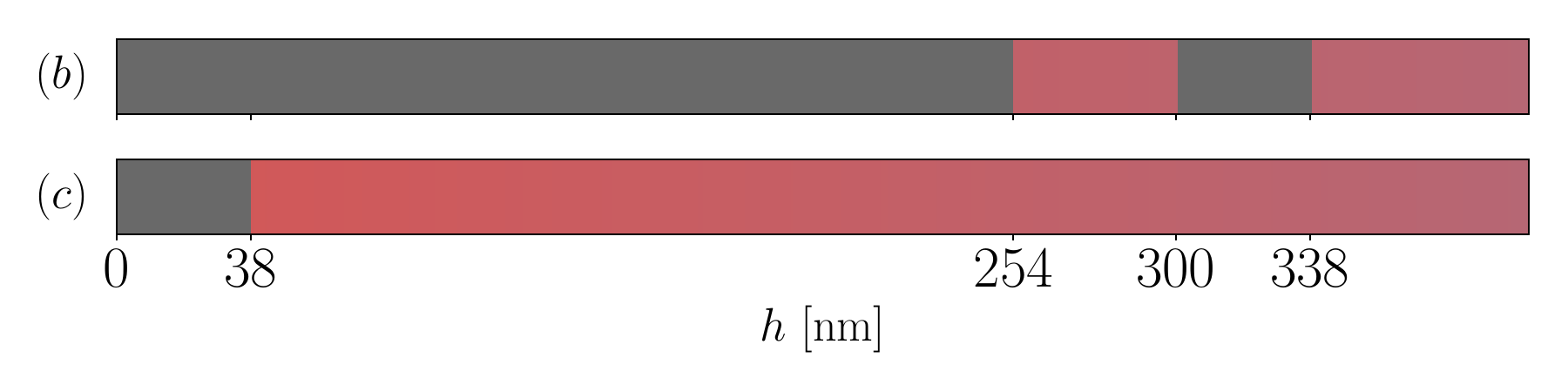} 
    \caption{$(a)$ Thickness range for each technique used associated with the corresponding errors $\epsilon$ determined from the theoretical spectrum of Eq.~\ref{eq:Intensityexpression}. $(b,c)$ Gray zones where the measurement of film thickness $h$ is not possible for our sensor operating in the range ($b$) $[450, 800]$~nm and for a hypothetical sensor operating in the range ($c$) $[100, 1000]$~nm. No grey zones zone observed beyond 350 nm.}\label{fig7}
\end{figure}
\end{center}

The previous section describes a systematic method for determining film thickness when at least two spectral minima are detected. Figure~\ref{fig7}~$(a)$ summarizes the thickness range for each technique used, along with the corresponding errors $\delta h/h$ determined from the theoretical spectra of Eq.~\ref{eq:Intensityexpression}.
Obviously, spectra with a low signal to noise ratio are unsuitable for such analysis. 
Furthermore, setting aside the specific case $p_d=1/2$ discussed in Section~\ref{subsec34}, when only a single extremum is present, the film thickness cannot be determined. 

Depending on the spectral range of the spectrometer, this occurs for different intervals of $\lambda$.
For example, Fig.~\ref{fig7}~($b$) illustrates the indeterminate thickness regions (gray zones) computed from Eq.~\ref{eq:Intensityexpression} for our spectral interval $[450, 800]$~nm. As an illustration, we also plot in Fig.~\ref{fig7}~($c$) the gray zones for a hypothetical sensor operating in the spectral range $[100, 1000]$~nm. 
In the broader $[100, 1000]$~nm range, any film thickness down to 38~nm can, in principle, be determined if the whole spectrum is known. 
In contrast, within the narrower spectral range $[450, 800]$~nm, two gray zones exist. Even with full knowledge of spectrum information, the intervals $[0, 254]$~nm and $[300, 338]$~nm remain inaccessible for film thickness determination.

The accuracy of the thickness estimation also depends on the refractive index $n(\lambda)$. 
For example, assuming a constant value of $n(\lambda) = 1.333$, as commonly done, can result in an error of up to 50 nm in the thickness determined using the FFT method.
We have summarized in Table \ref{tab:thickness_tableau} the effect of considering a constant optical index instead of using the expression of $n(\lambda)$ obtained by fitting the experimental data by the Cauchy's law. The deviation is close to or smaller than the error bars. 
\begin{table}[h!]
\centering
\begin{tabular}{lcc}
\hline
\textbf{Technique} & \textbf{With $n(\lambda)$} & \textbf{With $n = 1.333$} \\
\textbf{} & \textbf{$h \pm \delta h$ [nm]} & \textbf{$h \pm \delta h$ [nm]} \\
\hline
FFT (without padding) & $3400 \pm 800$ & $3500 \pm 800$ \\
Peak fitting & $1430 \pm 10$ & $1450 \pm 10$ \\
Scheludko & $495.7 \pm 0.2$ & $495.7 \pm 0.2$ \\
\hline
\end{tabular}
\caption{Effect of using a constant optical index on the thickness measurement for the three different methods detailed in Section 3.\label{tab:thickness_tableau}}
\end{table}

\section{Time-resolved analysis of soap films}\label{sec5}

\begin{center}
\begin{figure}[h]
    \centering
    \includegraphics[width=\columnwidth]{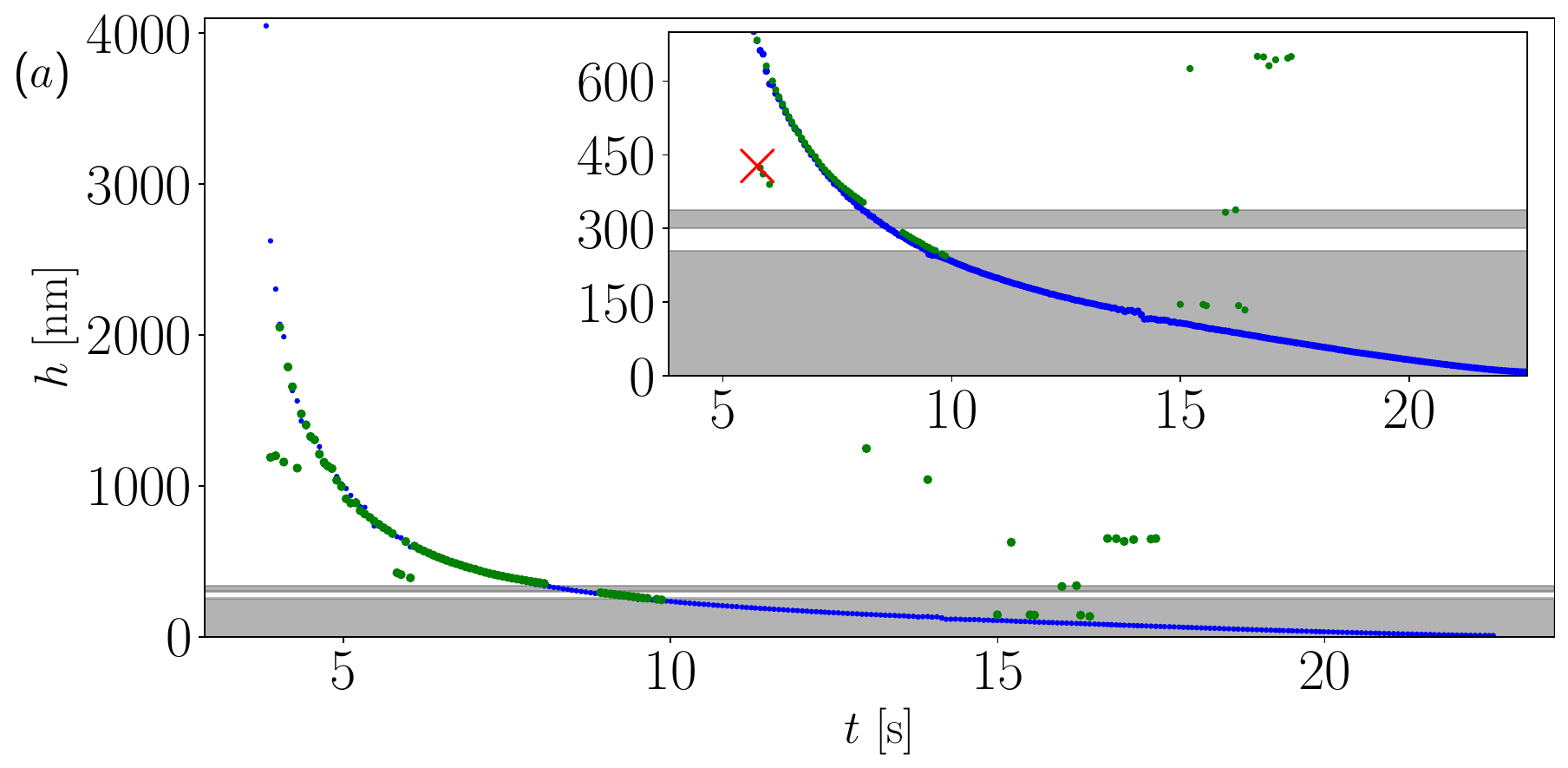} 
    \includegraphics[width=\columnwidth]{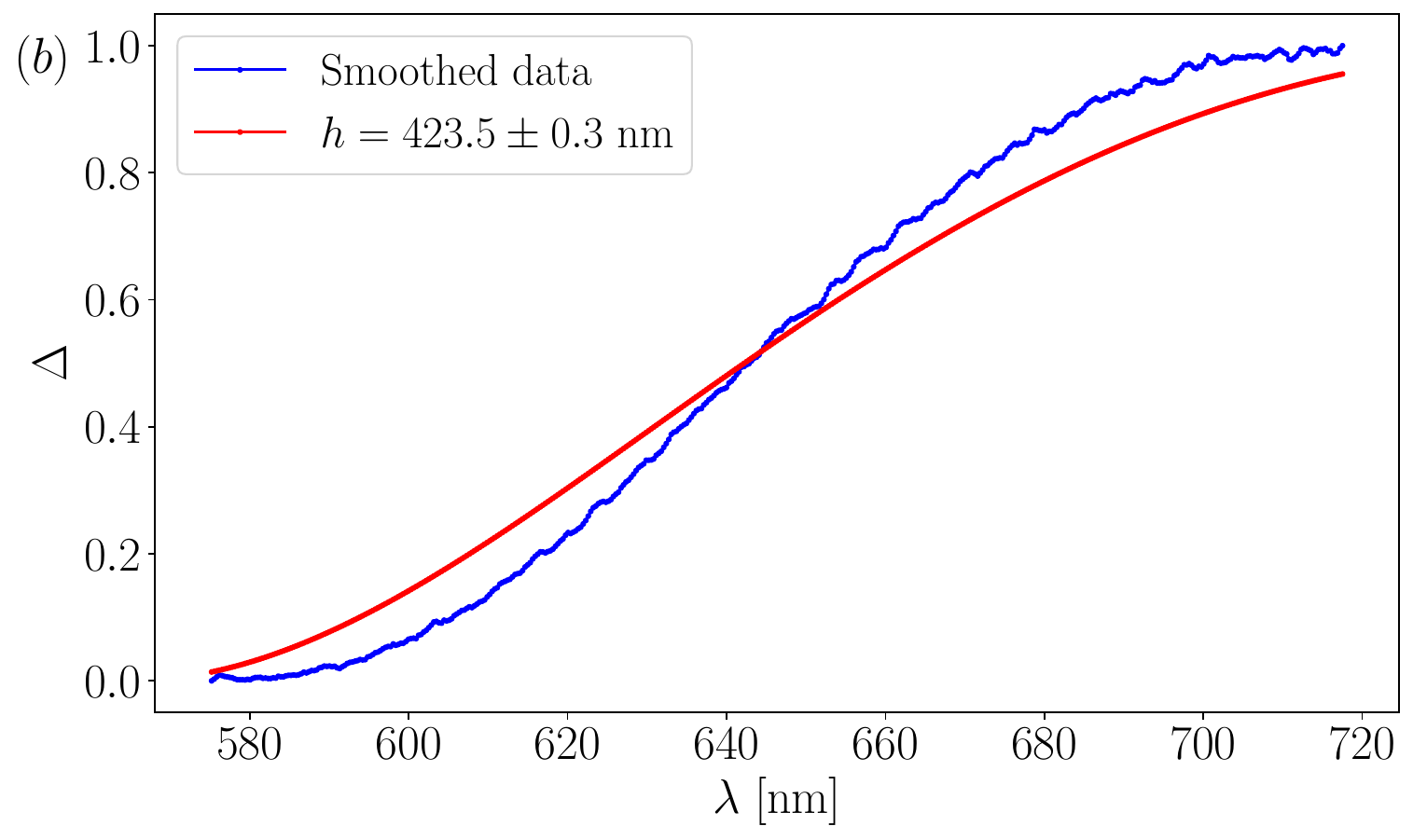} 
    \caption{$(a)$ Evolution of film thickness over time during elongation and thinning. Green dots correspond to thickness values obtained only using the method described in Section \ref{sec4}. Blue dots indicate thickness values reconstructed using the memory-based correction approach. Inset shows zoomed-in view of the $[0,\,700]$~nm region. $(b)$ $\Delta(\lambda)$ calculated for the spectrum indicated by the red cross \textcolor{red}{$\times$} in the inset of Fig.~8~$(a)$ compared to the function $\Delta(\lambda)$ calculated for the thickness value obtained with an incorrect interference order.}
    \label{fig8}
\end{figure}
\end{center}

Thanks to the different methods described in Section \ref{sec4}, it is possible to perform an analysis in time as the film thins. 
The results are presented in Fig.~\ref{fig8} (green dots).
Due to the limitations previously discussed, certain thickness ranges remain inaccessible. 
They are highlighted by gray horizontal zones. 
Inset in Fig.~\ref{fig8} shows a zoomed-in view of the $[0,\, 700]$~nm region. 
We will benefit from the dynamics of foam films, which are expected to thin continuously, to fill-in these gaps.
We call our method a memory-based analysis since it uses the thickness measured at time $t$ as an information to help measuring the one at time $t + \rm{d}t$.

After the film reaches a thickness of approximately $h=1 \, \mu$m, the thinning process becomes continuous and slow, with no abrupt transitions in the interference order (\textit{e.g.}, from order $m$ to order $m-2$) within the acquisition rate of the spectra.
Additionally, no film re-thickening is observed. 
This continuity enables manual correction of gaps and artifacts using a memory-based approach. 
The interference order can be reliably inferred throughout the entire curve step by step, storing the interference order for a given thickness to deduce that of the subsequent thickness, which will be identical or decremented.
The values of $I_\text{max}$ and $I_\text{min}$ can also be stored in memory for evaluating equations \ref{eq:DELTA_definition} and \ref{eq:h_expression} in order to compute $\Delta(\lambda)$ at any time $t$, even in the gray zones. 
Based on this memory approach, the manually corrected data are shown as blue dots in Fig.~\ref{fig7}. 
This memory-based method enables thin film measurements down to a few nanometers. 

The inset of Fig.~\ref{fig8} reveals a slight deviation between the corrected time-resolved data (blue dots) and the results obtained from the automatic method (green dots), particularly around a film thickness of approximately 350~nm. 
This difference arises from the way the maximum intensity values are used in the evaluation of Eq.~\ref{eq:h_expression}. 
In the manual time-resolved approach, a global maximum – corresponding to the peak intensity across all spectra for each interference order $p$ – is used uniformly for the entire dataset. 
In contrast, the automatic method relies on a local maximum, extracted individually from the spectrum recorded at each specific time $t$.

Note that some green points outside the grey zones are far away from the blue ones. This is because the program fails to identify the interference order $m$ correctly. To obtain the green points, we have performed a semi-automatic treatment, using the same parameters for peak detection consistently. This leads to incorrect peak detection and to incorrect identification of the interference order. In Fig.~\ref{fig8}~$(b)$, we plotted the experimental values of $\Delta(\lambda)$ together with the fit by Eq. \ref{eq:DELTA_definition} for the spectrum indicated by the red cross \textcolor{red}{$\times$} in the inset of Fig.~\ref{fig8}~$(a)$. It is obvious from such a fit that the value of the thickness is not accurate. If this occurs while treating the data, there can be different strategies to address the problem. One could adjust the parameters used for the peak detection or improve the experimental setup. We chose to use our memory based analysis because it is easily implementable in our particular case. Nevertheless, the thinning curve is accurately captured for most of its duration by the automatic method without any additional adjustments or the use of the memory-based analysis.

\section{Conclusion}\label{sec6}

In this article, we describe a systematic procedure for determining the thickness of a thin film using interferometry. The procedure is illustrated by three typical spectral scenarii in range $[450,\,800]$~nm obtained from TTAB soap films and implemented in the \textit{optifik} python library.

The first step of the procedure consists in counting the number of peaks, their number can be a hint to choose the best of the three methods that we describe in the article.

In the first method, where the spectrum exhibits numerous peaks, a Fast Fourier Transform (FFT) is applied to extract the main optical interference pattern, which is directly related to the film thickness $h$. 

In the second method, where the spectrum contains fewer interference peaks, the FFT method becomes unreliable. Instead, we determine the thickness by assigning interference orders through a fit or a RANSAC analysis on the quantity $n(\lambda)/\lambda$ at peak positions. Although this approach requires at least two detectable peaks, it introduces greater uncertainty in the estimated thickness $h$ if few peaks are detected. We fixed 5 peaks as the lower boundary.

The third method involves spectra with two to four interference peaks. In this case, the film thickness is determined by identifying the correct interference order of the two rightmost peaks using Eq.~\ref{eq:h_expression}, evaluated across several candidate orders $m$. 

The limitations of the method are discussed in Section~\ref{sec4}. As the technique is inherently interferometric, spectral quality is paramount; low SNR spectra reduce reliability. 
The spectral range also determines the measurable thickness domain, with broader ranges (\textit{e.g.}, [100,\,1000]~nm vs. [450,\,800]~nm) allowing access to wider thickness intervals. 
Furthermore, inaccuracies in the wavelength-dependent refractive index $n(\lambda)$ may significantly affect thickness determination accuracy. 
When a single interference maximum is observed, corresponding theoretically to an interference order $p = 1/2$,  the baseline intensity $I_{\text{min}}$ measured in the absence of film can be used in Scheludko renormalization, allowing thickness measurements in the range [84,\,150] nm.

Finally, a time-resolved analysis of TTAB films undergoing thinning is presented. When the film thickness continuously decreases below 1~$\mu$m without abrupt changes in interference order, a memory effect enables manual interpolation of data gaps. A comparison is made between the direct application of the method and an adjusted approach that compensates for artifacts and fills data gaps.

This interferometric technique is versatile and applicable to any non-opaque thin film, regardless of whether the film is thickening, thinning, or remaining stable over time. 
Together with the article, we provide a python library, \textit{optifik}, in which the initial peak detection as well as the three different methods are coded. Note that fully automated use remains delicate. 
It requires spectra with good contrast, and some adaptation should be made by the user, who can also take advantage from the additional experimental information, as we do here using the known dynamics of a draining soap film to fill the gaps.

This method can be adapted for transmission interferometry. In such a case, the calculations presented in Section~\ref{sec2}  must be modified, starting with the expression for the intensity $I_r$.

\section*{Acknowledgements}

We are very grateful to Alice Etienne-Simonetti, Marina Pasquet, Jonas Miguet, Lorène Champougny, and Laurie Saulnier for their previous work on developing the setup and the initial codes. Funding by the National Research Agency (ANR - 22-CE06-0029) is acknowledged.

\section*{Author contribution statement} 

The study was conceived and designed by ER, AS, FB, and VZ. Experimental execution and subsequent data analysis were carried out by VZ. VZ and FB were responsible for the implementation of the analysis in the open-source Python library Optifik, for testing and writting the documentation. The initial manuscript was drafted by VZ, and all authors (VZ, ER, AS, FB) contributed critically to its revision and approved the final version. 
 
\section*{Data availibility statements} The source code and the dataset are provided in \cite{optifik_github}.

\section*{Declarations}

The authors declare no conflict of interest.

\bibliographystyle{unsrt}

\begin{thebibliography}
\bibitem{whittaker_xviiifunctions_1915}
E.~T. Whittaker.
\newblock {On} the {Functions} which are represented by the {Expansions} of the
  {Interpolation}-{Theory}.
\newblock {\em Proceedings of the Royal Society of Edinburgh}, 35:181--194,
  1915.
\end{thebibliography}

\end{document}